\documentclass[groupedaddress,showpacs,showkeys,amssymb,twocolumn,aps]{revtex4-1}
\usepackage[pdftex]{graphicx}
\usepackage{amsmath} 
\usepackage{epsf}                                                                                             
\usepackage{color}                   

\usepackage{mathtools} % for "\mathclap" macro
\usepackage{amsfonts}
\usepackage{amssymb}
\usepackage{accents}

\usepackage{verbatim}
\newcommand{\be}{\begin{equation}}                                                                            
\newcommand{\ee}{\end{equation}}

\begin{document}                                                                                              
                                                                                                              
\title{Dual symmetry in a generalized Maxwell theory}

\author{F. T. Brandt}
\email{fbrandt@usp.br}
\author{J. Frenkel}
\email{jfrenkel@if.usp.br}
\affiliation{Instituto de F\'{\i}sica, Universidade de S\~ao Paulo, S\~ao Paulo, SP 05508-090, Brazil}
\author{D. G. C. McKeon}
\email{dgmckeo2@uwo.ca}
\affiliation{
Department of Applied Mathematics, The University of Western Ontario, London, ON N6A 5B7, Canada}
\affiliation{Department of Mathematics and Computer Science, Algoma University,
Sault St.Marie, ON P6A 2G4, Canada}

\begin{abstract}                                                                                            
We examine Podolsky's electrodynamics, which is noninvariant 
under the usual duality transformation.
We deduce a generalization of Hodge's star duality, which leads to a dual gauge field and restores 
to a certain extent the dual symmetry. The model becomes fully dual symmetric asymptotically, when it reduces 
to the Maxwell theory. We argue that this strict dual symmetry directly implies the existence of the basic 
invariants of the electromagnetic fields. 
\end{abstract}

\date{\today}


\maketitle                     

%\date{\today}

\section{Introduction}

Many symmetries that led to our understanding of quantum field theories, have their origin in the Maxwell 
theory describing the electromagnetic interactions. One such symmetry 
is duality which, in its simplest form, is the invariance of the source-free Maxwell equations under the 
interchange of electric and magnetic fields. Although the physical meaning of 
the dual symmetry is not yet entirely clear, there has been much work on the implications 
of this symmetry \cite{r1}. Apart from the related duality in supersymmetric and superstring theories \cite{r2,r3,r4,r5},
there has also been some interest in the extension of the dual symmetry to 
the classical Yang-Mills theory \cite{r6}. In this non-Abelian gauge theory, the Hodge dual field ${}^\star F_{\alpha\beta}$,
in contrast to the gauge field $F_{\alpha\beta}$, is not derivable from a potential, so that the 
usual dual symmetry is in general absent. 

The main purpose of this work is to study this problem in the context of a simpler (Abelian) gauge theory,
where the issues associated with the lack of a manifest dual symmetry may be easier to analyze and understand. 
One such model is Podolsky's generalized Maxwell theory \cite{r7}, which preserves the linear character of the field equations, 
yet avoids the divergences present in the Maxwell theory at short distances. 
Due to this behaviour, certain difficulties of classical electromagnetism, like the $4/3$ problem or non-causal 
effects are naturally solved in the context of Podolsky's electrodynamics \cite{r8,r9} . 
At large distances compared with some quantum scale, this model effectively reduces to the Maxwell theory. 
(For more recent studies of this theory see, for example, references \cite{r10,r10a,r11,r11a,r12,r12a}.) 

In Podolsky's model, a strict dual symmetry is also absent, which is apparent in the fact that the 
usual Hodge dual field is not generally a gauge field. We derive a generalized Hodge star duality, which 
reduces to the usual star operation at large distances. This leads to a dual gauge field derivable from 
a potential and restores, to a certain extent, the dual symmetry. This behaviour may reproduce, in a simpler context,
some analogous features which occur in the Yang-Mills theory. Thus, the present analysis may be helpful to understand a possible 
mechanism for the restoration of a dual symmetry in non-Abelian gauge theories. 

In a source-free region, Podolsky electrodynamics exhibits asymptotically, at large distances compared with the electron 
Compton wavelength, a complete dual symmetry. Since in this case the model practically reduces to the Maxwell theory,
it may be worth to examine some implications of the strict dual symmetry. 
%To this end, we examine 
We thereby deduce the electromagnetic invariants of the source-free Maxwell equations,
without an explicit resort to the Lorentz transformations properties of the fields. Using just the linearity and 
locality of the transformations, we show that the full duality directly implies the existence of the Lorentz invariants 
$E^2 - B^2$ and $\vec E\cdot\vec B$.

\section{Duality in Podolsky electrodynamics}
This theory  \cite{r7} may be described, in the source free case, by the action
\begin{equation}\label{e1}
S(l) = -\frac{1}{16\pi}\int d^4x\left[
F^{\alpha\beta} F_{\alpha\beta} + 2 l^2(\partial_\beta F^{\alpha\beta}) (\partial^\gamma F_{\alpha\gamma}) 
\right]   ,
\end{equation}
where $F_{\alpha\beta} = \partial_\alpha A_\beta - \partial_\beta A_\alpha$
is the electromagnetic gauge field tensor %and $l$ is a parameter with dimension of length. 
and $l$ is a length scale induced by quantum processes, which
regularises the classical Maxwell theory at short distances \cite{r8,r9}.   %xxxxx
%
%The cutoff $l$ which regularises the classical theory, may be of the same 
%order as the {\it effective}  radius of the electron charge distribution \cite{r8}. 
%It has been shown \cite{r13} that such a radius, of order of the electron Compton
%wavelength, may be effectively induced by processes which occur in quantum electrodynamics.
%In the limit $l\rightarrow 0$, this model clearly reduces to Maxwell theory.

The Euler-Lagrange equations following from \eqref{e1} are given by
\begin{equation}\label{e2}
(1-l^2\Box)\partial^\alpha F_{\alpha\beta}  = 0
\end{equation}
which leads to the generalized Gauss and Amp{\`e}re equations
\begin{equation}\label{e3}
(1-l^2\Box) \nabla\cdot\vec E = 0;   \;\;\; 
(1-l^2\Box) \left[\nabla\times\vec B -\frac{1}{c}\frac{\partial\vec E}{\partial t}\right] = 0 .
\end{equation}
By virtue  of the above form of the electromagnetic field tensor $F_{\alpha\beta}$, the Hodge dual tensor
\begin{equation}\label{e4}
{}^\star F_{\alpha\beta} =  \frac{1}{2} \epsilon_{\alpha\beta\mu\nu} F^{\mu\nu}
\end{equation}
satisfies the Bianchi identity
\begin{equation}\label{e5}
\partial^\alpha {}^\star F_{\alpha\beta}  = 0 .
\end{equation}
Since ${}^\star F_{\alpha\beta}$
can be obtained from $F_{\alpha\beta}$ by interchanging $\vec E\rightarrow \vec B$ and $\vec B\rightarrow -\vec E$,
\eqref{e5} is equivalent to the homogeneous equations
\begin{equation}\label{e6}
\nabla\cdot\vec B = 0;   \;\;\; 
\left[\nabla\times\vec E + \frac{1}{c}\frac{\partial\vec B}{\partial t}\right] = 0 .
\end{equation}
One can see from \eqref{e3} and \eqref{e6} that Podolsky's theory is no longer dual invariant in general.
%({\color{red} Does that mean that quantum corrections in normal QED breaks the dual symmetry?})
%under the above interchange of electric and magnetic fields. 
%However, since the equation \eqref{e3} effectively reduces \cite{r7}
%at distances much bigger than $l$, to the corresponding Maxwell equations, the dual
%symmetry is asymptotically recovered at large distances.  
We will next argue that there exists some potential
$\underaccent{\tilde}{A}_\mu$, which satisfies the equation
\begin{equation}\label{e7}
(1-l^2\Box) \; {}^\star F_{\alpha\beta}  =  \partial_\alpha \underaccent{\tilde}{A}_\beta - \partial_\beta \underaccent{\tilde}{A}_\alpha
\end{equation}
so that \eqref{e2} may be interpreted as the Bianchi identity for 
$(1-l^2\Box) \; {}^\star F_{\alpha\beta}$. To this end, we note that \eqref{e7} implies the relation
\begin{equation}\label{e8}
(1-l^2\Box) \; \left[\partial_\rho {}^\star F_{\alpha\beta} + 
\partial_\alpha {}^\star F_{\beta\rho} + \partial_\beta {}^\star F_{\rho\alpha} \right]  =   0.
\end{equation}
Due to the fact that the dual operation is reflexive, ${}^\star({}^\star F_{\alpha\beta}) = -F_{\alpha\beta}$, 
\eqref{e8} leads to the Eq. \eqref{e2}:
\begin{equation}\label{e9}
-(1-l^2\Box)\; \partial^\alpha{}^\star({}^\star F_{\alpha\beta})   = (1-l^2\Box)\partial^\alpha  F_{\alpha\beta} = 0.   
\end{equation}
In order to determine explicitly the potential $\underaccent{\tilde}{A}_\mu$,
we use an alternative method \cite{r6}, where $F_{\alpha\beta}$
are  considered as field variables, subject to the constraint \eqref{e5}. 
By imposing this condition, one can remove the inherent redundancies comprised in the set of variables
$\{F_{\alpha\beta}\}$. This procedure can be enforced by introducing Lagrange multipliers $\Lambda_\mu$ and constructing the action
\begin{eqnarray}\label{e10}
\bar S(l) &=& -\frac{1}{16\pi}\int d^4x\left[
F^{\alpha\beta} F_{\alpha\beta} \right. 
\\ \nonumber 
&+& \left. 
 2 l^2(\partial_\beta F^{\alpha\beta}) (\partial^\gamma F_{\alpha\gamma})   - 4 \Lambda^\beta \partial^\alpha {}^\star F_{\alpha\beta}
\right]   ,
\end{eqnarray}
Extremizing \eqref{e10} with respect to $F_{\alpha\beta}$ and $\Lambda^\beta$ we then get, apart from \eqref{e5}, the equation
\begin{eqnarray}\label{e11}
&F_{\alpha\beta} - l^2 \partial^\rho (\partial_\beta F_{\alpha\rho} - \partial_\alpha F_{\beta\rho}) = 
\\\nonumber 
&-\frac{1}{2} \epsilon_{\alpha\beta\mu\nu}(\partial^\mu\Lambda^\nu - \partial^\nu\Lambda^\mu).
\end{eqnarray}
Expressing the electromagnetic tensors in terms of their duals, 
we obtain from \eqref{e11}, after some algebra which makes use of the constraint \eqref{e5},  
the relation
\begin{equation}\label{e12}
(1-l^2\Box) \; {}^\star F_{\alpha\beta}  =  \partial_\alpha \Lambda_\beta - \partial_\beta \Lambda_\alpha.
\end{equation}
Comparing \eqref{e7} and \eqref{e12}, we see that the potential $\underaccent{\tilde}{A}_\mu$ is just the Lagrange multiplier $\Lambda_\mu$.

We note here that the general solution of \eqref{e7} can be written in the form
\begin{equation}\label{e13}
{}^\star F_{\alpha\beta}  =  {\cal F}_{\alpha\beta} +
(1-l^2\Box)^{-1} (\partial_\alpha \underaccent{\tilde}{A}_\beta - \partial_\beta \underaccent{\tilde}{A}_\alpha)
\end{equation}
where the operator $(1-l^2\Box)^{-1}$ may be interpreted, for slowly varying fields,  as the series 
$1 + l^2 \Box + \dots$ and ${\cal F}_{\alpha\beta}$ is a solution  of the homogeneous  equation
\begin{equation}\label{e14}
(1-l^2\Box) {\cal F}_{\alpha\beta}(x)  =  0 .
\end{equation}
It turns out that there exist non-trivial solutions of this equation. % \cite{r7,r8}. 
For example, in the static case one finds that
\begin{equation}\label{e15}
{\cal F}_{\alpha\beta}(\vec x)  =  \nabla^2 \int d^3x^\prime \frac{J_{\alpha\beta}(\vec x^\prime)}{|\vec x -\vec x^\prime|}
\;{\rm exp}\left(-\frac{|\vec x-\vec x^\prime|}{l}\right) 
\end{equation}
where $J_{\alpha\beta}$ is an external source  which vanishes at points $\vec x$ inside the source free region. 
We can see from \eqref{e15}, that ${\cal F}_{\alpha\beta}(\vec x)$                       
is exponentially vanishing at distances much larger than $l$, or equivalently 
in the limit $l\rightarrow 0$, as expected from \eqref{e14}. 
                                                                              
From the above equations, it follows that, in contrast to the gauge field  $F_{\alpha\beta} = \partial_\alpha A_\beta - \partial_\beta A_\alpha$, 
the dual field ${}^\star F_{\alpha\beta}$ is generally not derivable from a potential. 
However, we can see from \eqref{e7} that one may generalize the Hodge star duality as
\begin{equation}\label{e16}
\underaccent{\tilde}{F}_{\alpha\beta} \equiv (1-l^2\Box) {}^\star F_{\alpha\beta}  = \frac 1 2 \epsilon_{\alpha\beta\mu\nu} (1-l^2\Box) F^{\mu\nu} ,
\end{equation}
so that the modified dual field $\underaccent{\tilde}{F}_{\alpha\beta}$ is also a gauge field derivable 
from the potential $\underaccent{\tilde}{A}_\mu$: $\underaccent{\tilde}{F}_{\alpha\beta} = \partial_\alpha \underaccent{\tilde}{A}_\beta - \partial_\beta \underaccent{\tilde}{A}_\alpha$.
Such a  field satisfies, from  \eqref{e5}, the equation
\begin{equation}\label{e17}
(1-l^2\Box)^{-1} \partial^\alpha \underaccent{\tilde}{F}_{\alpha\beta} =  0.
\end{equation}
%The general solution \cite{r7} of Eq. \eqref{e2} includes, in particular, the class of solutions of an equation of the form \eqref{e17}.             
%Thus, within this  restricted class, this theory exhibits a generalized dual symmetry. 
It may be verified that the solutions of equations \eqref{e2} and \eqref{e17} 
are interchangeable, up to terms of order ${\rm exp}(-|\vec x|/l)$  which become
%differ by a term of order , which becomes 
small at distances larger than $l$. Thus, the dual gauge field
$\underaccent{\tilde}{F}_{\alpha\beta}$ restores to this extent the dual symmetry.
A complete dual symmetry is obtained in the asymptotic region where the parameter $l$
may be effectively set equal to zero, in which case Podolsky electrodynamics reduces to the Maxwell theory.

\section{Lorentz invariants from dual symmetry}
We will show next that the full dual symmetry of the source-free Maxwell equations directly leads
to the existence of the electromagnetic invariants. We recall that in a relativistic covariant theory, the 
electric and magnetic fields transform as a Lorentz tensor, so that the electromagnetic fields in two
inertial frames $K^\prime$  and $K$ satisfy the conditions \cite{r14,r15}
\begin{equation}\label{e18}
E^\prime{}^2 - B^\prime{}^2 = E^2 - B^2 
\;\;\; \mbox{and}  \;\; \vec E^\prime \cdot \vec B^\prime = \vec E \cdot \vec B. 
\end{equation}
In order to derive these invariants from duality, let us consider a general transformation which relates
the fields in  $K^\prime$ and $K$. 
In a homogeneous and isotropic space-time, which is assumed by the relativity principle, the
transformation should be linear.  
Moreover, such a transformation cannot contain derivatives of the fields, since it must be local. 
The locality ensures that the values of the fields at some space-time point in $K$, uniquely determine the fields observed in 
$K^\prime$ at the same space-time point. Thus, a general transformation which relates $\vec E^\prime$ to $\vec E$ and $\vec B$ must have the form
\begin{eqnarray}\label{e19}
\vec E^\prime &=& a_1\vec E + a_1^\prime \vec B +b_1\vec\beta\times\vec B +b^\prime_1\vec\beta\times\vec E 
\\\nonumber &&
+ d_1 (\vec\beta\cdot\vec E) \vec\beta + d^\prime_1 (\vec\beta\cdot\vec B) \vec\beta 
\end{eqnarray}
where $c \vec\beta$ is the velocity of  $K^\prime$ relative to $K$ and the dimensionless coefficients  
$a_1$, $a^\prime_1$, $b_1$, $b_1^\prime$, $d_1$, $d_1^\prime$                                                   
are real functions of   $\beta^2$. 
The six structures in \eqref{e19} may be independent, since transformations in 4-dimensional space-time may be characterized 
by six parameters, corresponding for example to three rotation angles and three Lorentz boosts.  
Using the dual symmetry under $\vec E\rightarrow \vec B$ ($E^\prime\rightarrow \vec B^\prime$) and 
$\vec B\rightarrow -\vec E$ ($B^\prime\rightarrow -\vec E^\prime$),
it follows from \eqref{e19} that the corresponding transformation of the magnetic field should have the form
\begin{eqnarray}\label{e20}
\vec B^\prime &=& a_1\vec B - a_1^\prime \vec E -b_1\vec\beta\times\vec E +b^\prime_1\vec\beta\times\vec B 
\\\nonumber &&
+ d_1 (\vec\beta\cdot\vec B) \vec\beta - d^\prime_1 (\vec\beta\cdot\vec E) \vec\beta 
\end{eqnarray}
It is now convenient to introduce the complex electromagnetic vector field \cite{r1,r14}
\begin{equation}\label{e21}
\vec F = \vec E + i \vec B. 
\end{equation}
Then, it follows from \eqref{e19} and \eqref{e20} that the general linear transformation of the complex vector
field $\vec F$          can be written as
\begin{equation}\label{e22}
\vec F^\prime = a \vec F - i b \vec\beta \times \vec F + d (\vec\beta\cdot\vec F)\vec\beta , 
\end{equation}
where %$a$, $b$ and $d$ are (dimensionless) complex coefficients given by: 
%\begin{equation}\label{e9}
$a = a_1 - i a_1^\prime$,  $b = b_1 + i b_1^\prime$ and $d = d_1 - i d_1^\prime$ 
are complex coefficients.
%\end{equation}
The inverse transformation can be obtained by changing     $\vec \beta\rightarrow -\vec\beta$   in \eqref{e22}:
\begin{equation}\label{e23}
\vec F = a \vec F^\prime + i b \vec\beta \times \vec F^\prime + d (\vec \beta\cdot\vec F^\prime)\vec\beta . 
\end{equation}
Substituting     \eqref{e23}       in     \eqref{e22}, we obtain the consistency conditions
\begin{equation}\label{e24}
a^2 - b^2\beta^2 =1;\;\; a+ d\beta^2 = 1,
\end{equation}
where we used the fact that $a \rightarrow 1$ as $\beta\rightarrow 0$.

Let us consider the parallel and perpendicular components of  \eqref{e22}       with respect to    $\vec\beta$. Using          
\eqref{e24} we then obtain the relations
\begin{subequations}\label{e25}
\begin{eqnarray}\label{e25a}
\vec F_{\|}^\prime &=&(a+d\beta^2)\vec F_{\|} = \vec F_{\|} \\
\vec F_{\perp}^\prime &=& a \vec F_\perp - i b \vec\beta\times\vec F_\perp .
\end{eqnarray}
\end{subequations}
These correspond to a rotation around  $\vec\beta$     by a complex angle 
$i\theta$ where  $\theta = \tanh^{-1}(\beta b/a)$. Thus, we get 
\begin{equation}\label{e26}
\vec F_\perp^\prime\cdot \vec F_\perp^\prime = (a^2-b^2\beta^2) \vec F_\perp \cdot \vec F_\perp  =
\vec F_\perp\cdot \vec F_\perp .
\end{equation}
From (\ref{e25a}) and (\ref{e26}) it follows that $\vec F^\prime\cdot\vec F^\prime = \vec F\cdot\vec F$ 
and hence
\begin{equation}\label{e27}
{E^\prime}^2 - {B^\prime}^2 + 2 i \vec E^\prime \cdot \vec B^\prime  =
{E}^2 - {B}^2 + 2 i \vec E \cdot \vec B  
\end{equation}
which leads to the Lorentz invariants (\ref{e18}). 
We have thus shown that the electromagnetic
invariants can be deduced from the dual symmetry of the source-free Maxwell equations.

%\section{Conclusion}
We finally remark that a photon mass would break the dual symmetry  
of the Maxwell equations.
Such a mass may be generated, for example, by the Stueckelberg mechanism \cite{r16,r17}
which preserves gauge invariance and renormalizability. 
Hence, the  fact that  the photon remains massless could be explained
if duality would hold at a fundamental level.
(Measurements of the galactic vector potential place an
upper bound on the photon mass of $10^{-18}$ eV/c${}^2$ \cite{r18}.) 
This is an open problem which might be worthy of further examination.

\section*{Acknowledgments}
We thank CNPq (Brazil) for a grant and Prof. J. C. Taylor for helpful discussions.

\end{document}